\documentclass[%
 reprint,
 amsmath,amssymb,
 aps,
]{revtex4-2}
\usepackage{amsmath,amssymb}
\usepackage{color}
\usepackage{graphicx}
\usepackage{dcolumn}
\usepackage{bm}

\begin{document}

\preprint{APS/123-QED}

\title{Avalanche shapes in fiber bundle model}

\author{Narendra Kumar Bodaballa}
 \affiliation{Department of Physics, SRM University - AP, Andhra Pradesh 522 240, India}

\author{Soumyajyoti Biswas}
\email{soumyajyoti.b@srmap.edu.in}
\affiliation{Department of Physics, SRM University - AP, Andhra Pradesh 522 240, India}%

\author{Parongama Sen}%
 \email{psphy@caluniv.ac.in}
\affiliation{Department of Physics, University of Calcutta, Kolkata 700 009, India
}%

\date{\today}

\begin{abstract}
We study the temporal evolution of avalanches in the fiber bundle model of disordered solids, when the model is gradually driven towards the critical breakdown point. We use two types of loading protocols: (i) the quasi-static loading, and (ii) loading by a discrete amount. In the quasi-static loading, where the load is increased by the minimum amount needed to initiate an avalanche, the temporal shapes of avalanches are asymmetric away from the critical point and become symmetric as the critical point is approached. A measure of asymmetry follows a universal form $A\sim (\sigma-\sigma_c)^{\theta}$, with $\theta\approx 0.25$, where $\sigma$ is the load per fiber and $\sigma_c$ is the critical load per fiber. This behavior is independent of the disorder present in the system in terms of the individual failure threshold values. Thus it is possible to use this asymmetry measure as a precursor to imminent failure. For the case of discrete loading, the load is always increased by a fixed amount. The dynamics of the model in this case can be solved in the mean field limit. It shows that the avalanche shapes always remain asymmetric. We also present a variable range load sharing version of this case, where the results remain qualitatively similar.  
\end{abstract}

\maketitle


\section{\label{sec:level1}Introduction}
The intermittent avalanche dynamics are seen in a myriad of systems, starting from invasion of fluid in a porous media, moving domain walls in impure magnetic materials, compressed rocks prior to breakdown to sliding tectonic plates causing earthquakes. The statistical nature of these avalanches, particularly its scale free size and duration distributions, reveal information regarding the system that are near-universal in nature i.e., they depend on only a few parameters such as the interaction range in such systems, their dimensions are so on \cite{biswas2015statistical, pradhan2010failure, alava2006statistical}. 

The nature of the avalanches are also known for its use in hazard assessments on certain systems, where such assessments are crucial. For example, it is well established that the sizes of the avalanches, if sampled near a catastrophic failure point, show an exponent value that is smaller than what one would obtain if all avalanches are sampled \cite{scholz1968frequency}. Specifically, in shear dynamics of compressed granular media, the avalanche size distribution exponent is known to depend on the differential stress \cite{hatano2015common}. The same is observed in earthquake statistics i.e., the Guttenberg-Richter exponent changes in the regions where larger events are likely to occur \cite{schorlemmer2005variations, narteau2009common}. There are other ways to extract meaningful information from the avalanche statistics near the breaking point, for example the inequality of avalanches show universal precursors that signal imminent breakdown \cite{PhysRevE.106.025003, PhysRevE.108.014103}. 

In this work, we study the temporal variation of the breaking dynamics, i.e., the temporal shapes of avalanches when a system is driven slowly towards its catastrophic breakdown point. Particularly, we use the fiber bundle model, which is a paradigmatic model for fracture of disordered solids. When external load is applied gradually, the system goes through intermittent avalanches that, on average, grow in size and eventually culminate at the breaking point. However, when an avalanche is initiated by increasing the external load by a small amount, on an otherwise stable system, the avalanche continues through multiples steps of stress field readjustments. In each such step of readjustment, some local breaking events happen. The sum over all such events until the next stable point is called one avalanche. A time-scale separation is always assumed between the internal stress readjustments and the external loading rate, which makes it possible to consider the external load as a constant during an ongoing avalanche. Therefore, it is also possible to look at the temporal shape of the avalanches, measured in terms of the numbers of breaking events within a single avalanche in different steps of stress-field readjustments. 

The temporal shape of an avalanche have been looked at in avalanching systems before, both theoretically and experimentally \cite{laurson2013evolution, Gleeson2017,batool2022temporal}. Particularly, when an `elastic' interface is driven through a disordered medium, it goes through stick-slip motions, showing avalanches. Such situations are ubiquitous in various contexts such as invading fluid fronts, magnetic domain walls, mode-I fracture front and so on. With a `slow' (in terms of the time scale separation discussed above) drive and associated dissipation (acoustic and other such forms of energy emissions), dynamics of such driven interfaces reach a self-organized critical (SOC) state. Among other things, one property of an SOC system is that the critical point is an attractive fixed point and the SOC systems always reside close to it. With this being said, the temporal shape of the avalanches were shown to depend on the nature of `elasticity' of the driven interface. Particularly, it was shown \cite{laurson2013evolution} that for sufficiently local range of the interaction kernel in the driven interface, the time-reversal symmetry of the avalanches are broken, while for sufficiently non-local kernels the shapes are symmetric under time-reversal. In the mean field limit, therefore, the direction of time cannot be determined by looking at the data of stress field readjustments within an ongoing avalanche (or similar other manifestations of the dynamics). However, these assertions (dependence of time-reversal asymmetry on non-locality of interaction kernels) are studied only close to the critical point. 

Here we look at the variations of the avalanche shapes in disordered systems, starting far away from the critical point and gradually approaching the breakdown point. We find that the avalanche shapes are asymmetric away from the critical point and become symmetric as the critical point is approached, even for a mean field interaction. Given that the system is mean-field, the symmetry at the critical point is expected. For the same reason, therefore, a reduction of asymmetry could be an useful indicator of the approaching criticality. We show that there is a universal trend in approaching a time-reversal symmetric avalanche pattern in the fiber bundle model of fracture with quasi-static load increase that does not depend on the type of disorder present in the model. Finally, we also study other modes of load increase viz., the load increase by a discrete amount. Analytical solution of the dynamical equations show that the avalanche shapes under this loading condition is always asymmetric.

\begin{figure}
\centering
\includegraphics[width=8.5cm]{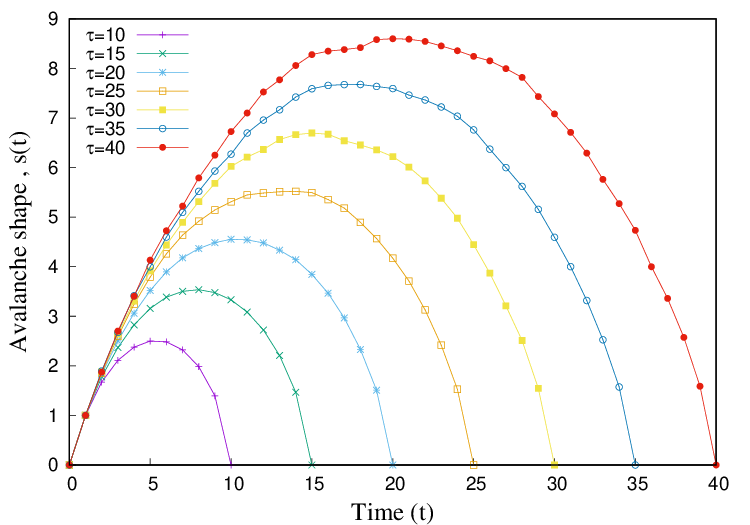}
\caption{Avalanche shape , s(t) for different specific duration of the avalanches with uniform threshold distribution under quasi static loading}
\label{time_shape_unscaled}
\end{figure}
\begin{figure*}
\centering
\includegraphics[width = 18 cm]{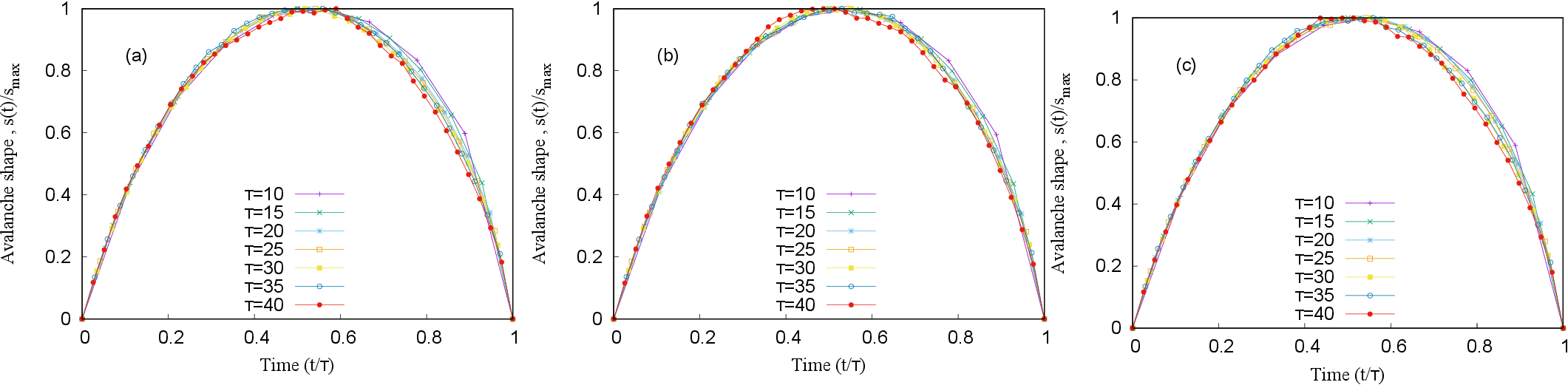}
\caption{Average avalanche shapes for avalanches with of duration $\tau$ for various disorder disorder distribution functions (a) Uniform distribution, (b) triangular distribution, (c) Gaussian distribution. The load redistribution steps $t$ are scaled by the avalanche duration $\tau$ and the sub-avalanche sizes (number of fibers breaking in one load redistribution) are scaled by the maximum value for each duration. While generally the shape is that of an inverted parabola, there are deviations from this symmetric shape, as the the system approaches failure (higher avalanche duration here is a proxy to higher applied load, which in-turn indicates imminent failure.}

\label{time_shape}
\end{figure*}
\begin{figure*}
\centering
\includegraphics[height=6.2cm]{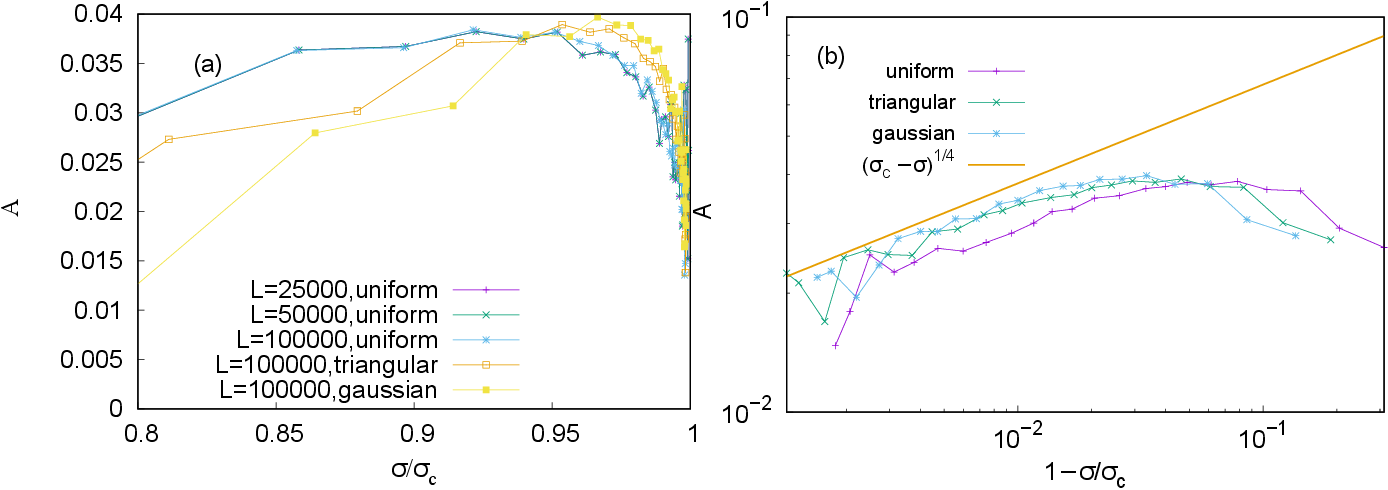}
\caption{Plot showing the variation of the asymmetry of the avalanches  (measured using Eq. (\ref{asymm_measure})) with the scaled external load on the system for which the avalanche of a particular duration had started. The larger duration avalanches start with a higher load. The initial small-sized avalanches are symmetric, but the avalanches become more asymmetric from higher loads. However, as the system nears the failure point, the avalanches start becoming more symmetric, which is then an indication of imminent failure.}
\label{asymm_load}
\end{figure*}

\begin{figure}
\centering
\includegraphics[width=8cm]{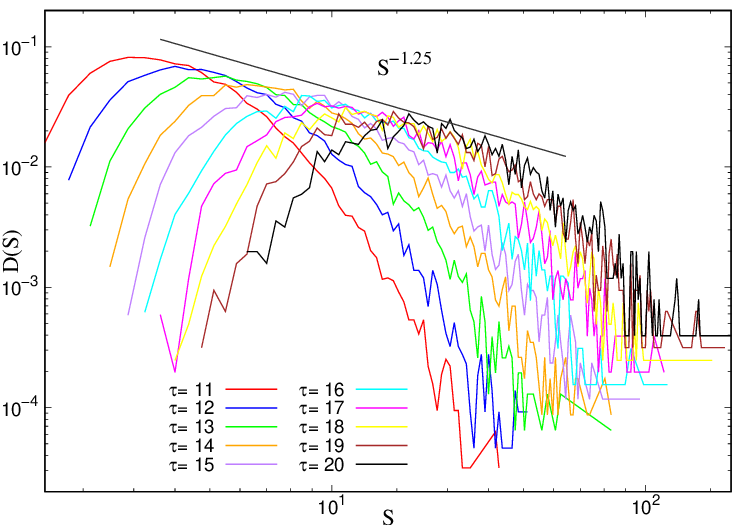}
\caption{Plot showing the probability of avalanche size D(S) for uniform distribution thresholds for different specific duration of avalanche $\tau$}
\label{probability_specific_time}
\end{figure}

\section{\label{sec:level2}The model and its avalanche dynamics}
The fiber bundle model is a threshold activated model for fracture of disordered solids that can capture many qualitative features of fracture dynamics, for example the intermittent scale-free avalanche statistics. 
In this study of evolution  \cite{laurson2013evolution,Gleeson2017} of fiber bundle with time, we consider  $N$  fibers arranged in parallel. These fibers carry an external mechanical load. The fibers are assumed to be linear elastic elements which can bear load up to a threshold \begin {math}{\sigma}_{th} \end {math}. After reaching the threshold capacity, the fibers break irreversibly.  The values of these thresholds for the individual fibers are randomly sampled from a probability density function \begin {math}{\rho}({\sigma}_{th}) \end {math}. Unless otherwise specified, this function is uniform in the domain (0,1) \cite{PhysRevE.67.046122}. 

To initiate the dynamics of the failure process, an initial load \begin {math}{\sigma}_0 \end {math} is applied on each fiber \begin{math}{\sigma}_i\end{math}.  The fibers with threshold $\sigma_{th}>\sigma_i$ will survive and the remaining fibers will break irreversibly. The  load applied on the failed fibers  is redistributed equally to all the surviving fibers. Because of this increase in the load of the surviving fibers, further fibers may break. The process of redistribution will continues until no more fibers break i.e., all thresholds are higher than the applied load per fiber value or the entire system has broken down.
In the former case, the load is increased  to restart the dynamics.


The number of fibers broken between two successive load increase is called an avalanche $S$, while the number of fibers breaking in each step of redistribution is called a sub-avalanche $s$. We denote by $t$ the time step of load redistribution within an avalanche ($t$ running from 0 to $\tau$ would give an avalanche of duration $\tau$), and with $t^{\prime}$ the number of load increments. So, if there is an avalanche of duration $\tau$ at time $t^{\prime}$, then the corresponding sub-avalanche time variable would run from $t=0$ to $\tau$, in steps of $1/\tau$ in terms of $t^{\prime}$.

There are two kinds of loading protocol that we will deal with here. These are distinguished based on how the load is increased once the system reaches a stable configuration. One protocol is to increase the load (uniformly on all surviving fibers) by an amount $\Delta \sigma$ that is the minimum required value in order to break the fiber which is closest to its breaking point i.e., \begin {math}{\delta}{\sigma}=min({\sigma}_{th}^{i} - {\sigma}_i) \forall  \; i  \end{math} where \begin{math} i \end{math} denotes the intact fibers. This process will restart the dynamics by breaking exactly one fiber (the one closest to the failure threshold) and the avalanche may continue. This is called the quasi-static loading protocol. In the other protocol, the increment of load is always by a constant amount to the system (note that increment of load per fiber will progressively increase, since the number of surviving fibers decreases). This is called the discrete loading protocol. The avalanche initiation step here will typically cause more than one fiber to break, and the dynamics may continue thereafter. These two types of loading protocols are profoundly different in terms of the corresponding avalanche statistics. Particularly, while both protocols produce scale-free size distributions for the avalanches, the exponent values differ (-2.5 in the former 
[Eq. (208) of \cite{pradhan2010failure}] and -3 in the latter [Eq. (139) of  \cite{pradhan2010failure}]. 


In this work, we study the avalanche shapes for both of these loading protocols under mean field and localized load sharing and outline a recipe to detect imminent catastrophic failure.

\section{\label{sec:level3}Average avalanche shape in the case of quasi-static loading}
As mentioned above, with quasi-static loading, the avalanche of fibre failures \cite{batool2022temporal}  starts with a single fiber failing (the one closest to failure), causing the load to be redistributed onto the remaining intact fibres, which can lead to further breaking events.  The size of avalanche $S(\tau)$ with duration $\tau$ contributed by several sub-avalanches $s_t$ are related as 

\begin{equation}  
S = \sum_{t=0}^{\tau} s_t
\label{ava_def}
\end{equation}
where the index $t$ refers to an integer values that tracks the number of redistribution steps or sub-avalanches.
The duration of an avalanche $\tau$ is a proxy to the total load on the system i.e., the duration is a monotonically increasing function of the total load on the system on average. In calculating the asymmetry of the avalanches, therefore, we first collect all avalanches of a particular duration in the evolution of a particular sample and also that in many ensembles of nominally similar (having the same failure threshold distribution) samples. The average shape, defined as
\begin{equation}
    \langle s_t(\tau) \rangle=\frac{1}{E}\sum_{\alpha=1}^{E} s_t^{(\alpha)} (\tau),
\end{equation}
where $E$ is the number of ensembles,
of all those avalanches of a particular duration is then calculated (see Fig. 2). Then, the x-axis is scaled by the selected duration (making it vary between 0 to 1) and the y-axis is scaled by the maximum height of the avalanche shape profile (making it also varying between 0 to 1). This process was repeated for different types of threshold distributions: uniform, Gaussian and triangular each within the range (0,1) (see Fig. \ref{time_shape}). 
Then, the asymmetry of the resulting average avalanche profile (denoted by $\langle s_t\rangle)$ is calculated from \cite{hohmann2019enforced}

\begin{multline}
A( \langle s(\tau) \rangle ) =\frac{1}{\langle s_{max}(\tau)\rangle \tau} \sum_{t=0}^{\tau}  | \langle s_{t}(\tau) \rangle -\\ min \{\langle s_{t}(\tau)\rangle , \langle s_{\tau-t}(\tau)\rangle
\} |
\label{asymm_measure}
\end{multline}
where,$\langle s_t(\tau) \rangle=\frac{1}{E}\sum_{\alpha=1}^{E} s_t^{(\alpha)} (\tau)$.  Defined in this way, a fully (time-reversal) symmetric shape would give $A=0$. As can be seen in Fig. \ref{asymm_load} , a plot of the above mentioned asymmetry measure with the corresponding average initiating load (load per fiber at the time when an avalanche was started) shows a shape decrease as the critical load ($\sigma_c$, the load per fiber value at which the system collapses) is approached (see Fig. \ref{asymm_load} (a)). This form of the decrease is found to be universal i.e., it does not depend on the particular threshold distribution used (see Fig. \ref{asymm_load}(b)). Indeed, an approximate power-law fitting suggests $A \sim (\sigma_c-\sigma)^{1/4}$ for all three types of threshold distributions. Furthermore, these results also do not seem to have any systematic system size dependence.  

It is interesting to look at the size distribution of the avalanches of a particular duration (see Fig. \ref{probability_specific_time}). While the size distribution of all avalanches is expected to be a power-law (with exponent value -5/2), when looked at for a specific duration, the size distributions are no longer scale-free. Indeed, the scale of the avalanche sizes (for example, the average or the most probable value) increases monotonically with duration. This is expected for the correspondence between avalanches of longer duration and that of the average initiating load on the system. A fit of the most probable values show a power-law of exponent -1.25. This non scale-free structure of the individual components of an overall scale-free distribution has a resemblance of similar observations in wealth distribution models \cite{PATRIARCA2006723}.  


\section{\label{sec:level4}Solutions to the dynamical process in discrete loading}
The breaking dynamics (the evolution of $U(t)$) can be solved analytically for the case of discrete loading. While such solutions have been previously looked at for the case when the system is very close to the braking point \cite{PhysRevE.67.046122}, here we attempt a solution for the entire range of the dynamics.



To do that, we need to consider the initial threshold distribution to be uniform within the range (0,1). Then, for any subsequent step of the dynamics, the threshold distribution of the remaining part of the system will 
 uniform distribution within domain $[{\sigma}_L,{\sigma}_R]$. As the dynamics progresses, the values of $\sigma_L$ will change (increase) as more and more weaker fibers are eliminated from the system. In what follows, we first solve $U(t)$ for fixed values of $\sigma_L$ and $\sigma_R$ and then numerically track the evolution of $\sigma_L$ to find the recursive dynamics of $U(t)$. 

The cumulative distribution of the thresholds can be written as
\begin{equation} \label{eqn}
P({\sigma}_{th}) = 
     \begin{cases}
       0 & {\sigma}_{th}<{\sigma}_L\\
       \dfrac{{\sigma}_{th}-{\sigma}_L}{{\sigma}_R-{\sigma}_L} & {\sigma}_L\leq{\sigma}_{th}\leq {\sigma}_R\\
       1 &{\sigma}_{th}\ge {\sigma}_R    
     \end{cases}
\end{equation}

For the above mentioned threshold distribution of the stress values, the  recursive relation of failure process written as
\begin{equation}
\frac{dU}{dt}=\frac{1}{{\sigma}_R-{\sigma}_L}\frac{{\sigma}_R U-{\sigma}-U^2({\sigma}_R - {\sigma}_L)}{U}
\end{equation}
then, using the partial fraction method above equation reads
\begin{multline}
- \frac{t}{{\sigma}_R-{\sigma}_L}=\frac{1}{2} \int\frac{2U-{\sigma}_R}{U^2-({\sigma}_R -{\sigma}_L) -{\sigma}_R U+{\sigma}_o} dU  \\  +  \frac{1}{2} \int\frac{{\sigma}_R}{U^2-({\sigma}_R -{\sigma}_L) -{\sigma}_R U+{\sigma}_o} dU
\end{multline}.
Now setting the $b=\frac{{\sigma}_R}{{\sigma}_R-{\sigma}_L}$ ,$c=\frac{{\sigma}_o}{{\sigma}_R-{\sigma}_L}$,$z=\frac{2U-b}{2\big(\frac{b^2}{4}-c)}$, the transformed equation after solving will be
\begin{multline}
-\frac{t}{{\sigma}_R -{\sigma}_L}=\frac{1}{2}\log \big|U^2({\sigma}_R-{\sigma}_L)-{\sigma}_R U + {\sigma}_o | \\ +\frac{b}{4\sqrt{\frac{b^2}{4}-c}}\log\bigg|\frac{y-1}{y+1}\bigg| + \log k
 \end{multline}
 Again writing the equation in its original variable form, solution gives us the evolution of the surviving fraction of fibers $U_t$ when the system is applied with an initial load per fiber ${\sigma}_o$
\begin{multline}\label{eq:11}
-\frac{t}{{\sigma}_R -{\sigma}_L} = \frac{1}{2}\log \big|U^2({\sigma}_R-{\sigma}_L)-{\sigma}_R U + {\sigma}_o | \\ +  \frac{b}{4\sqrt{\frac{b^2}{4}-c}} \log \bigg|\frac{2U-b-2\sqrt{\frac{b^2}{4}-c}}{2U-b+2\sqrt{\frac{b^2}{4}-c}}\bigg| + \log \;k
\end{multline}
where $\log\;k =-\frac{1}{2}\log \big|U^2({\sigma}_R-{\sigma}_L)-{\sigma}_R  + {\sigma}_o | \\ +  \frac{b}{4\sqrt{\frac{b^2}{4}-c}} \log \bigg|\frac{2-b-2\sqrt{\frac{b^2}{4}-c}}{2-b+2\sqrt{\frac{b^2}{4}-c}}\bigg|$. While a closed form solution of the above equation is not possible, the dynamics of the model exactly follow it i.e., the full range of the time evolution of the FBM under a discrete loading for a uniform threshold distribution is solved here. This is a general case of the special condition where such dynamics were solved \cite{PhysRevE.67.046122} only near the stable points, leading to an evaluation of the relaxation time scale. 

However, here we are interested with the temporal shape of the avalanches, which can be obtained numerically from the above equation.

\subsection{Numerical calculation and simulations of failure process in ELS model}
\begin{figure}
\centering
\includegraphics[width=8.5cm]{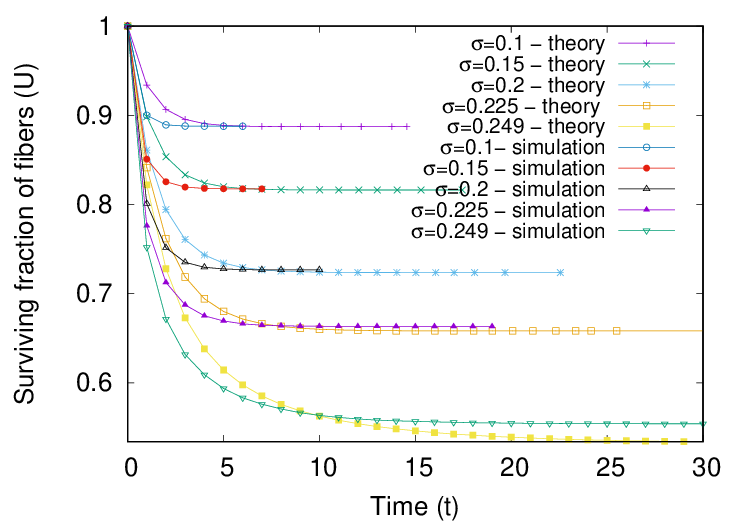}
\caption{The theoretical estimates  (numerically solving  Eq. (\ref{eq:11})) and the simulation  results of the surviving fraction of fibers $U(t,\sigma_0)$ as a function of redistribution (time) steps, when the system is loaded with several values of initial load per fiber ${\sigma}_0$, with initial condition $U(t=0,\sigma_0)=1$  in each case. The threshold distributions are uniform in (0,1) and the system size $L=100000$ for the simulations.}
\label{combined_U_one}
\end{figure}

The temporal profile of the surviving fraction of fibers $U_t$ with the initial load ${\sigma}_0$, can be obtained by numerically solving Eq.  (\ref{eq:11}).
The results are shown in Fig. \ref{combined_U_one} where simulations for $N=100000$ fibers are also shown for comparison. 
This result can then be used to calculate the variation of the surviving fraction, when a fixed amount of load increment is made successively on the system, as soon as it reaches a stable configuration. This we call the discrete loading mechanism, as mentioned above.

\begin{figure}
\centering
\includegraphics[width=8.5cm]{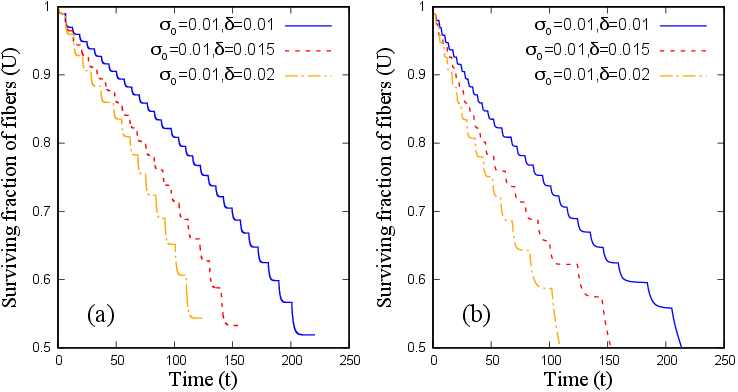}
\caption{The theoretical (a) (from Eq \ref{eq:11}) and simulation (b) results for which disorder in system is retained by uniform distribution with system size of $N=100000$ for the recursive dynamics of the surviving fraction of fibers $U(t)$ with discrete load increase. Both figures show steps decay of the surviving fraction, each step indication a load increment. The difference in the curves arise because of the error added due to the numerical calculation of  the solution obtained for the surviving fraction of fiber}
\label{U_recurssive}
\end{figure}

This dynamics following a single step loading is necessary to calculate  the dynamics of the system with discrete loading as follows: 
  To investigate the dynamics of successive loading steps, Eq. (\ref{eq:11}) is numerically solved for various combinations of $ {\sigma}_L, {\sigma}_R,{\sigma}_o ,U$ . This allows for the calculation of the temporal evolution of the avalanche shape during the failure process. To solve Eq. (\ref{eq:11}), the value of  $U$  is iteratively decreased by 0.00001, starting from $U=1$. The corresponding values of $t$ are computed until the solution to the calculated value of time begins to decrease, which are considered physically irrelevant solutions. Once this occurs, ${\sigma}_o$ is adjusted to increase such that it takes the value ${\sigma}_o/U^* + {\delta}$ and ${ \sigma } _ L$ is adjusted to the value ${\sigma}_o/U^*$.  Again Eq. (\ref{eq:11}) is then solved with $U=1$, decreasing it by 0.00001 for a new combination of $  {\sigma}_L, {\sigma}_R,{\sigma}_0,U$. This process of numerical solution is continued until the loading step, where the system completely collapses. Here, $U$ in each loading step is normalized  to the previous discrete loading's final surviving fraction of fiber. This normalization ensures that the analysis takes into account the relative survival of the fiber in comparison to the previous loading condition. The numerical solutions and simulation results are shown in Fig. \ref{U_recurssive}.

\begin{figure}
\centering
\includegraphics[width=8.5 cm]{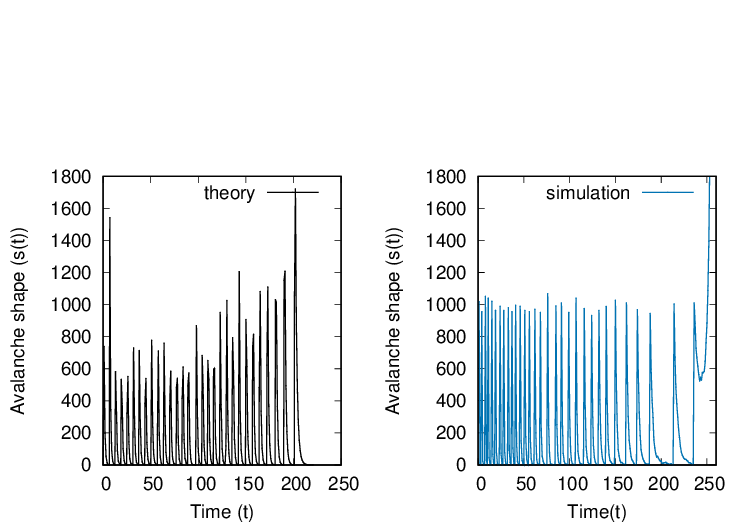}
\caption{Theoretical(a)  and simulation (b) (as in Fig. \ref{U_recurssive} results of the recursive dynamics for the avalanche size with system size (N=100000 fibers) (differentiation of Eq. (\ref{eq:11})). The area between two successive zeros of the function give one avalanche of size $S$ .}
\label{S_recursive}
\end{figure}

  The variation of the surviving fraction of fibers  $U$  of discrete loading  is shown in Fig. \ref{U_recurssive}. We have also numerically calculated (from Eq. (\ref{eq:11})), the evolution of the avalanche sizes by taking the theoretical estimate data from the Fig. \ref{U_recurssive} and calculated the avalanche shape with difference of surviving fraction of fibers for the corresponding time redistribution steps which is shown in Fig .\ref{S_recursive} (a). These results are then compared with simulation of fiber bundle with $N=100000$ and the shape of the avalanches are shown in Fig. \ref{S_recursive}(b), which is qualitatively similar to the results shown in Fig. \ref{S_recursive}(a) . However, there is a quantitative difference which arises because of the choice of decrement value of  $U$ while numerically solving the Eq. (\ref{eq:11}).
  
The avalanche shape can then be calculated by $s(t)={\Delta}U/{\Delta}t$. The avalanche shapes are expectantly asymmetric with the evolution of time as shown in  Fig. \ref{S_recursive}.

\subsection{Avalanche shape prior to catastrophic failure with variable load redistribution range}

here, we explore the behavior of a  fiber bundle model subjected to discrete loading. I when a fiber fails, its load is distributed equally to the nearest R intact fibers, following the LLS (local load sharing) scheme. By adjusting the interaction range parameter R i.e., distance between the intact fiber to the failed fiber, we calculate the shape of the avalanche just before a catastrophic event using Monte Carlo simulations. The simulations are conducted on a system consisting of $50,000$ fibers.

During averaging of the simulation, we record the sub avalanche sizes of the last three avalanches that occur just before the catastrophic event. We then calculate the average shape of these three avalanches to understand the overall pattern before the event takes place. It's worth noting that the avalanche shape exhibits asymmetry concerning the time leading prior to the catastrophic event under discrete loading conditions. Additionally, we observe that the peak of the avalanche shape gradually diminishes as the loading process progresses, which is influenced by the increment of the applied load {$\delta$}.

\begin{figure*}
\centering
\includegraphics[scale=1.5]{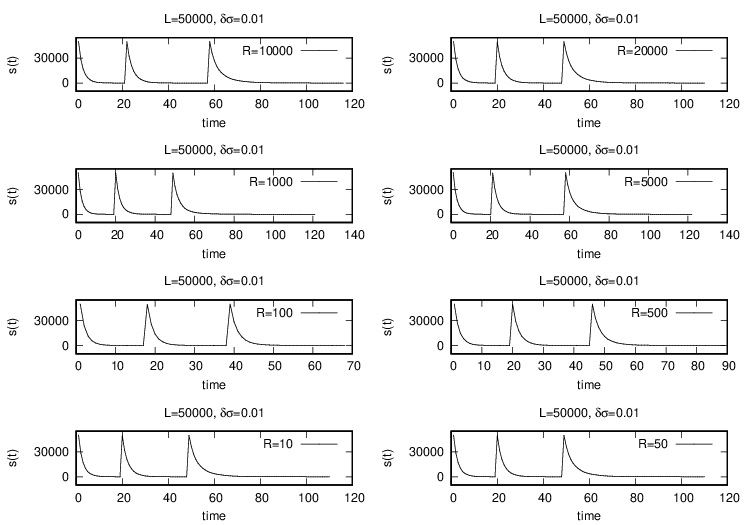}
\caption{simulation results showing last 3 avalanches before the catastrophic event under discrete loading, load redistribution is varied by different interaction ranges $R$}
\label{R_avl_shape}
\end{figure*}                                                                       
\section{Discussions and Conclusions}
The intermittent avalanche dynamics in driven disordered systems show remarkable statistical regularities. Particularly, the size distribution of avalanches are known to follow power-law distributions, with exponent values that are only dependent on the ranges of interactions, dimensionality of the system and do not depend on the other system specific details, such as the distributions of disorders within the system and so on. Such regularities have been studied in great details, both theoretically and experimentally. However, much less is known about the time variation profile of the individuals avalanches. 

Here we looked at this question of temporal avalanche shapes in the fiber bundle model of fracture, primarily in the mean-field limit. While it is known that near the critical point, the avalanche shapes, for mean-field systems, are symmetric with respect to time-reversal, here we have shown that away from the critical point, under quasi-static loading protocol, the symmetry is broken. Therefore, a quantification of asymmetry for the avalanches leads to a drastic drop as the system approaches the failure point (see Fig. \ref{asymm_load}).
The functional form of the variation of the asymmetry in the avalanche shapes are universal, in the sense that it does not depend upon the threshold distributions of the fibers. Also, there is no systematic system size dependence.
Therefore, monitoring the shapes of avalanche could serve as a useful indicator of the imminent failure. 

We also study the avalanche shapes for discrete loading protocol. While the dynamical evolution of the avalanche shapes are partially analytically tractable in this case, the shapes are always asymmetric, both in the mean field and local load sharing limit.  

In conclusion, the temporal shapes of avalanche for the fiber bundle model under quasi-static loading tends from an asymmetric to symmetric shape, in a universal manner, as the catastrophic breakdown point is approached. 

\begin{acknowledgments}
The numerical simulations are performed in HPCC Surya at SRM University - AP.
PS acknowledges financial support from SERB project MTR/2020/000356.
\end{acknowledgments}


\bibliography{apssamp}

\end{document}